\documentclass{jfm}
\usepackage{graphicx}
\usepackage{amsmath}
\usepackage{epstopdf, epsfig}
\usepackage{placeins}
\usepackage{amsmath}

\newcommand{\blu}[1]{\textcolor{black}{#1}}
\usepackage{color}

\title{Aspect-ratio dependent heat transport by baroclinic acoustic streaming}

\author{Jacques Abdul Massih\aff{1},
    Remil Mushthaq\aff{2},
  Guillaume Michel\aff{3}\corresp{\email{guillaume.michel@sorbonne-universite.fr}}
 \and Gregory P. Chini\aff{1,2}}

\affiliation{
\aff{1}Program in Integrated Applied Mathematics, University of New Hampshire, Durham, NH 03824, USA
\aff{2}Department of Mechanical Engineering, University of New Hampshire, Durham, NH 03824, USA
\aff{3} Sorbonne Université, CNRS, Institut Jean Le Rond d’Alembert, F-75005 Paris, France}

\begin{document}
\maketitle

\begin{abstract}
Standing acoustic waves have been known to generate Eulerian time-mean `streaming' flows at least since the seminal investigation of Lord Rayleigh in the 1880s. Nevertheless, a recent body of numerical and experimental evidence has shown that inhomogeneities in the ambient density distribution lead to much faster flows than arise in classical Rayleigh streaming. The emergence of these unusually strong flows creates new opportunities to enhance heat transfer in systems in which convective cooling cannot otherwise be easily achieved. To assess this possibility, a theoretical study of acoustic streaming in an ideal gas confined in a rectangular channel with top and bottom walls maintained at fixed but differing temperatures is performed. A two time-scale system of equations is utilized to efficiently capture the coupling between the fast acoustic waves and the slowly evolving streaming flow, enabling strongly nonlinear regimes to be accessed. A large suite of numerical simulations is carried out to probe the streaming dynamics, to highlight the critical role played by baroclinically-generated wave vorticity and to quantify the additional heat flux induced by the standing acoustic wave. Proper treatment of the two-way coupling between the waves and mean flow is found to be essential for convergence to a self-consistent steady-state, and the variation of the resulting acoustically-enhanced steady-state heat flux with both the amplitude of the acoustic wave and the $O(1)$ aspect ratio of the channel is documented. For certain parameters, heat fluxes almost two orders of magnitude larger than those realizable by conduction alone can be attained.

\end{abstract}


\section{Introduction}
\label{sec:intro}

Standing acoustic waves in homogeneous fluids have long been known to generate Eulerian mean flows. In the 19th century, the observation of air flows in a Kundt's tube by \cite{DVORAK_1876} was one of the three phenomena that led \cite{RAYLEIGH_1884} to develop his pioneering analysis of acoustic streaming. A general framework, reviewed by \cite{RILEY_2001}, has since been achieved: the Reynolds stress divergence induced by a standing acoustic wave generates a streaming flow provided this stress divergence cannot be balanced by a mean pressure gradient; that is, provided the acoustic wave has non-zero vorticity. If acoustic wave attenuation takes place over a much longer time scale than the acoustic period, fluctuating vorticity is localized in thin oscillatory boundary layers, and the characteristic streaming velocity scale $U_{s_*}=U_*^2/a_*$, with $U_*$ the maximum acoustic-wave velocity and $a_*$ the speed of sound. This streaming velocity remains modest and limits practical applications to microfluidics, where acoustic forcing is used to mix dilute chemicals along the direction transverse to the micro-channel \citep{Bengtsson_2004}. Recently. it has been shown that larger streaming velocities can be achieved if the solid boundaries include sharp edges having radii of curvature smaller than the thickness of the oscillating boundary layer, a promising finding albeit one that is difficult to realize experimentally \citep{Huang_2013, Zhang_2019}.

Standing acoustic waves in \emph{inhomogeneous} fluids drive streaming flows having radically different features. This essential outcome of the last two decades stems from experimental evidence presented by \cite{Loh_2002}, \cite{HYUN_2005} and \cite{Stockwald2014HighlyEM}, who reported streaming velocities in stratified gases two orders of magnitude larger than the estimate $U_*^2/a_*$ and streaming patterns qualitatively different from that predicted by Rayleigh's theory \citep{Dreeben_2011}. This change of phenomenology is also apparent in direct numerical simulations (DNS) of the compressible Navier-Stokes equations \citep{LIN_2008, AKTAS_2010}. Based on these results, \cite{chini_2014} derived a new set of wave/mean-flow interaction equations that are able to capture the dynamics of streaming flows in an inhomogeneous gas. These authors demonstrated that the physical origin of the enhanced streaming flows is the baroclinic production of fluctuating vorticity, as is evident by taking the curl of the linearized Euler equation, \emph{viz.}
\begin{equation}
\nabla \times \left( \rho_0 \frac{\partial \mathbf{u}'}{\partial t} = - \nabla p' \right) \Rightarrow \frac{\partial (\nabla \times \mathbf{u}')}{\partial t} = \frac{(\nabla \rho_0) \times (\nabla p')}{\rho_0^2},\label{vorticity}
\end{equation}
where $\rho_0$ is the non-uniform background density, $\mathbf{u}'$ is the acoustic wave velocity, $p'$ is the acoustic wave pressure, $\nabla$ is the spatial gradient operator and $t$ is the time variable. Whereas the fluctuating vorticity required to drive streaming flows in homogeneous fluids results from viscous torques and is confined to thin Stokes boundary layers, in (stably) density stratified fluids wave vorticity can readily fill the entire domain owing to its generation via an inviscid process. Accounting for this fundamental change leads to a characteristic streaming velocity $U_{s_*}=U_*$ in strongly inhomogeneous fluids. Given that the \blu{acoustic Mach number} $U_*/a_*$ typically is very small compared to unity, this result provides a rationale for the large-amplitude streaming flows previously reported. \cite{Karlsen_2016} extended this procedure to inhomogeneities in compressibility (negligible in gases) and demonstrated that, in an inhomogeneous gas, the wave-induced Reynolds stress divergence (i.e. the acoustic force density $\mathbf{f}_\mathrm{ac}$) can be expressed as
\begin{equation}
\mathbf{f}_\mathrm{ac} = - \frac{1}{2} \:\overline{|\mathbf{u}'|^2} \:\nabla \rho_0,\label{RSD_Karlsen}
\end{equation} 
where the overbar indicates a time average. This concise representation confirms that the forcing is inviscid and associated with baroclinicity (also proportional to $\nabla \rho_0$, as shown in (\ref{vorticity})). Moreover, this expression correctly suggests that two-way coupling between the fast acoustic waves and the slowly-evolving streaming flow can be realized. Specifically, in the framework of \cite{chini_2014} and \cite{Karlsen_2016}, the waves drive a streaming flow that advects inhomogeneities in density that, in turn, feed back on the wave velocity field $\mathbf{u}'$. Furthermore, given that the velocity field of the standing acoustic wave depends on the background density distribution over the entire domain, this two-way coupling is spatially non-local.

In addition to manifesting more complex dynamics, this new regime of streaming is practically important. In high-intensity discharge lamps, a temperature difference of several thousand degrees exists between the arc and the tube wall, enabling baroclinic acoustic streaming to be used to improve the efficiency of these lamps dramatically \citep{Dreeben_2011, chini_2014}. Similarly, streaming flows in thermoacoustic devices strongly depend on the inhomogeneous temperature distribution \citep{daru_2021, Daru_2021b}. In microfluidics, properly accounting for inhomogeneities in both density and compressibility is crucial for obtaining accurate predictions of the acoustically-driven mixing of different fluids \blu{\citep{Karlsen_2018, Pothuri_2019, qui2021}}. Acoustic wave forcing also has been proposed as an experimental means to mimic and tune gravity in regions where $\nabla  (\overline{\mathbf{u}'^2}) $ is spatially uniform, $\mathbf{f}_\mathrm{ac}$ then being locally similar to gravity in the Boussinesq approximation up to a gradient term \citep{koulakis_putterman_2021, Koulakis_2023}, although we stress that this connection no longer holds once two-way coupling sets in. Perhaps the most promising application involves the use of acoustics to enhance the rate of heat transfer from heated objects immersed in a cooler fluid medium, especially in scenarios in which forced convection is difficult to establish or natural convection does not occur (e.g. as for cooling electronic components aboard spacecraft). 

Acoustic streaming in a straight channel with a temperature differential imposed between the top and bottom boundaries is a simple yet practical configuration in which to study the effects of inhomogeneity. Many prior studies have focused on the use of liquids as the working fluid, for which variations in compressibility also contribute to the acoustic force density \citep{Karlsen_2016}. 
These studies include the large set of numerical simulations performed by \cite{Kumar_2021} and \cite{RAJENDRAN2022} in which the effects of two-way coupling were neglected. Their results demonstrate that acoustics can either enhance or reduce the heat flux associated with natural convection depending on whether the acoustic wave displacements are, respectively, predominantly parallel or perpendicular to the channel walls. Here, we consider the second class of fluids, i.e. gases, typically air in experiments, approximated as an ideal gas in theoretical analyses. The experiments of \cite{NABAVI2008} clearly demonstrate that the streaming pattern derived in the homogeneous limit by \cite{RAYLEIGH_1884} is modified for temperature differences as low as a fraction of a degree. Results from DNS of the compressible Navier--Stokes equations for larger temperature differentials have been reported in \cite{LIN_2008}, \cite{AKTAS_2010} and \cite{Baran_2022}. The modest acoustic amplitudes considered in these simulations 
resulted in only minor modifications of the background density profile, allowing \cite{michel_chini_2019} to obtain an explicit expression for these baroclinic streaming flows by neglecting the two-way coupling. Larger acoustic amplitudes lead to two-way coupling, as analyzed theoretically by \cite{chini_2014}, \cite{Karlsen_2018} and \cite{michel_chini_2019} and realized experimentally by \cite{Michel_2021}. 

In the present study, the dynamics of an ideal gas undergoing standing acoustic-wave oscillations in a differentially-heated channel is investigated using an extension of the wave/mean-flow interaction equations derived in \cite{michel_chini_2019}. Our main objective is to characterise the strongly nonlinear regimes attained at large acoustic amplitudes that manifest two-way coupling. More specifically, we quantify the dependence of the acoustically-enhanced  heat flux on the geometry of the flow, i.e. on the aspect ratio $\delta = k_*H_*$, where $k_*$ is the acoustic-wave wavenumber and $H_*$ the height of the channel. Previous theoretical investigations have shown that the heat-transfer enhancement becomes negligible in the limit $\delta \rightarrow 0$ \citep{michel_chini_2019}, while recent DNS performed for modest acoustic amplitudes suggest that maximum enhancement is reached for $\delta = O(1)$, although these simulations did not achieve strong two-way coupling \citep{Malecha_2023}. Moreover, the scaling of the heat flux with the wave amplitude is largely undocumented in this regime. Presuming the background temperature field is homogenised in the interior of the domain as a result of the streaming-induced mixing, baroclinicity (and, hence, the acoustic force density) will localize near the walls. The consequence of this qualitative change of behavior on the generation of 
\blu{steady mechanical and thermal boundary layers} and on the overall heat transfer rate is also addressed in the present investigation. 

The remainder of this article is organised as follows. The two time-scale wave/mean-flow system governing the two-way coupled dynamics for arbitrary (but fixed) values of $\delta$ is introduced in \S \ref{sec:prob}. The numerical algorithm for simulating the wave/mean-flow interaction equations is described in \S \ref{sec:results}, where detailed results are also presented. A summary of our key findings and their potential implications is given in \S \ref{sec:conclusion}.

\section{Problem Formulation}
\label{sec:prob}
\subsection{Flow configuration}

\begin{table}
		\begin{center}
		\def~{\hphantom{0}}
    \begin{tabular}{lc}
	       \underline{Dimensional variable or parameter} & \underline{Definition}\\ 
        \\
	    $\tilde{\mathbf{u}}= (\tilde{u},\tilde{v})$ & Gas velocity\\	       
	    $\tilde{\rho}$ & Gas density\\
	    $\tilde{p}$ & Gas pressure\\
	    $ \tilde{T}$ & Gas temperature\\
	    $(\tilde{x},\tilde{y})$ & Horizontal, vertical (wall-normal) coordinate\\
	    $\tilde{t}$ & Time variable\\
	    $H_*$ & Channel height\\
	    \blu{$2\pi/k_*$} & \blu{Horizontal wavelength of acoustic wave}\\
	    $\mu_*$ & Dynamic viscosity\\
	    $\kappa_*$ & Thermal conductivity\\
	    $R_*$ & Specific gas constant\\
	    $(c_{v_*}, c_{p_*})$ & Constant volume, pressure specific heat coefficient\\
	    $a_*=\sqrt{(c_{p_*}/c_{v_*}) R_* T_*}$ & Background sound speed\\
	    $p_*$ & Background pressure\\
		\end{tabular}    
	\end{center}
	\caption{Dimensional variables and parameters.}
	\label{tab:dimensional_parameters}
\end{table}

The flow configuration is similar to that investigated by \cite{michel_chini_2019}. Dimensional variables are denoted with tildes and dimensional parameters with asterisks. 
These variables and parameters and their definitions are summarised in table~\ref{tab:dimensional_parameters}.

\begin{figure}
  \centerline{\includegraphics[width=0.7\linewidth]{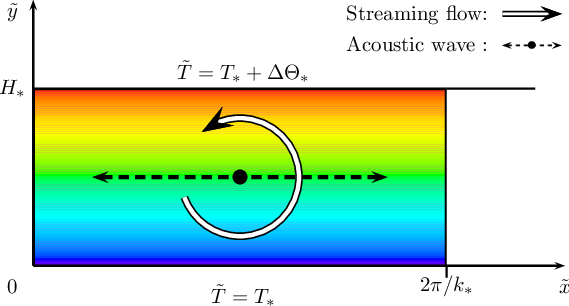}}
  \caption{Schematic of the two-dimensional system configuration, similar to \cite{michel_chini_2019}. An ideal gas is confined between two horizontal, no-slip and impermeable walls separated by a distance $H_*$. The temperatures of the lower and upper walls are fixed at $T_*$ and $T_* + \Delta \Theta_* $, respectively (but note that gravity is not included). A standing acoustic wave of horizontal wavenumber $k_*$ generates a counter-rotating cellular streaming flow spanning the channel that enhances the initially-diffusive heat flux.}
\label{fig1_setup}
\end{figure}

As illustrated in figure~\ref{fig1_setup}, we investigate the two-dimensional {\blu{(2D)}} dynamics of an ideal gas in an infinitely long channel of height $H_*$. No-slip and zero normal-flow boundary conditions on the velocity along with Dirichlet conditions on the temperature are imposed along each horizontal wall, with the temperature set to $T_*$ at $\tilde{y}=0$ and to $T_*+\Delta \Theta_*$ at $\tilde{y}=H_*$. This environment is assumed to be gravity-free and, thus, the imposed temperature differential $\Delta \Theta_*>0$ neither generates natural convection nor restoring forces. In the absence of acoustic forcing, a linear temperature profile is established owing to the strictly  diffusive heat flux across the channel. An unspecified external agency then drives a standing acoustic wave of angular frequency $\omega_*$ and wavenumber $k_*$ along the horizontal ($\tilde{x}$) direction to generate a streaming flow that enhances this heat flux. The horizontal \blu{(spatial) periodicity $2\pi/k_*$ of the acoustic wave} is presumed constant, being fixed in laboratory experiments, for example, by the finite horizontal length of the channel (or, more properly, cavity). Consequently, the angular frequency $\omega_*$ may vary with time as the temperature field slowly evolves. Moreover, \blu{the spatial phase of the acoustic wave is fixed by setting }
to zero the horizontal velocity component at $\tilde{x}=0$, i.e. $\tilde{u}(\tilde{x}=0, \tilde{y},\tilde{t})=0$, where $(\tilde{x},\tilde{y})$ are  the horizontal and vertical coordinates and $\tilde{t}$ is the time variable.

The flow is governed by the compressible Navier-Stokes equations, supplemented by the energy equation and the ideal gas equation of state,\\ 
\begin{equation} \label{D_momentum}
\tilde{\rho} \left[ \partial_{\tilde{t}} \tilde{\mathbf{u}} + ( \tilde{\mathbf{u}} \cdot \tilde{\mathbf{\nabla}} )\tilde{\mathbf{u}} \right] = - \tilde{\mathbf{\nabla}} \tilde{p}  +  \mu_* \left[ \tilde{\mathbf{\nabla}}^2 \tilde{\mathbf{u}} + \frac{1}{3}\tilde{\mathbf{\nabla}}  ( \tilde{\mathbf{\nabla}} \cdot \tilde{\mathbf{u}} ) \right],
\end{equation}

\begin{equation} \label{D_mass}
\partial_{\tilde{t}} \tilde{\rho} + \tilde{\mathbf{\nabla}}  \cdot \left( \tilde{\rho} \tilde{\mathbf{u}}  \right) = 0,
\end{equation}

\begin{equation} \label{D_heat}
\tilde{\rho} c_{v_*}  \left[ \partial_{\tilde{t}} \tilde{T} + ( \tilde{\mathbf{u}} \cdot \tilde{\mathbf{\nabla}} )\tilde{T} \right] = - \tilde{p} \left( \tilde{\mathbf{\nabla}} \cdot \tilde{u} \right)  + \kappa_* \tilde{\nabla}^2 \tilde{T},
\end{equation}

\begin{equation} \label{D_state}
\tilde{p} = \tilde{\rho} R_* \tilde{T},
\end{equation}
where $\tilde{\rho}$, $\tilde{p}$, $\tilde{T}$ and $\tilde{\mathbf{u}}=(\tilde{u},\tilde{v})$ respectively stand for the density, pressure, temperature and velocity fields, and $\tilde{\nabla} = (\partial_{\tilde{x}}, \partial_{\tilde{y}})$.  As discussed in \cite{michel_chini_2019}, bulk viscosity and viscous heating are neglected and the variation of dynamic viscosity with temperature is ignored in (\ref{D_momentum})--(\ref{D_state}). 

\subsection{Scaling and non-dimensionalization}
\label{Scaling_ND}
Dimensionless variables and parameters are introduced using the scalings given in table~\ref{tab:scalings}. The small parameter subsequently used in the asymptotic analysis is the \blu{acoustic Mach number $\epsilon = U_*/a_*$, also referred to as the inverse Strouhal number of the oscillating flow.} The dimensionless temperature differential $\Gamma = \Delta \Theta_* / T_*$ is taken to be fixed and order unity as $\epsilon\to 0$ to characterize a stratified gas in which baroclinicity plays a crucial role. This distinguished limit is similar to that adopted in \cite{chini_2014} and \cite{michel_chini_2019}. In contrast with these previous investigations, which focused on long thin channels for which $\delta = O (\sqrt{\epsilon}) $, in this work we consider acoustic waves with horizontal wavelengths comparable to the height of the channel, i.e. $\delta = O(1)$. \blu{Note that the Reynolds and P\'eclet numbers based on the acoustic-wave oscillatory velocity and wavelength also characterize the streaming flow, which has typical velocities of order $U_{s*}=U_*$ and develops structures having a spatial periodicity comparable to that of the waves.
(In \cite{chini_2014} and \cite{michel_chini_2019}, cross-channel, i.e. vertical, diffusion is enhanced owing to the thinness of the channel, resulting in a disparity between the acoustic and streaming Reynolds numbers.)}


It is instructive to compare these scalings to typical experimental values. Using the definitions given in table~\ref{tab:scalings}, the acoustic cavity considered by \cite{Michel_2021}, for instance, would be characterized by the following parameter values: $\epsilon = 1.8\times 10^{-4}$, $\delta = 1.5$, $\Gamma \in [0.1,0.3]$, $Re=134$ \blu{, $Pe=86$ and $A \in [0.2,2]$ (where the wave amplitude $A = O(1)$ is introduced in \S \ref{sim_equations})}. 
Thus, the scaling requirements $\epsilon \ll 1$, $\delta = O(1)$, $\Gamma = O(1)$, $Re = O(1)$ and $Pe = O(1)$ encompass this setup.

	\begin{table}
		\begin{center}
		\def~{\hphantom{0}}
    \begin{tabular}{l c|  c  c c}
	       \underline{Variable} &   \underline{Scale} &   \underline{Parameter} &  \underline{Definition} &   \underline{Scaling}\\ 
	   & & & &\\
     $x$ & $k_*^{-1}$ & 
     \blu{Acoustic Mach} number $\epsilon$ & $U_*/a_*$  &$\epsilon \ll 1$\\
	    $y$ & $H_*$& Aspect ratio $\delta$ & $k_* H_*$ &$\delta = O(1)$\\
	    $t$ & $(a_* k_*)^{-1}$ & Temperature gradient $\Gamma$ & $\Delta \Theta_* / T_*$  & $\Gamma = O(1)$\\
	    $u$ & $a_*$ & Reynolds number $Re$ & $\rho_* U_*/ (k_* \mu_*)$ & $Re = O(1)$\\
	    $v$ & $(k_*H_*)a_*$ & P\'eclet number $Pe$ & $\rho_* c_{p_*} U_* / (k_* \kappa_*)$  & $Pe=  O(1)$\\
	    $\rho$ & $\rho_* \equiv p_* /(R_* T_*)$ & Specific heat ratio $\gamma$ & $c_{p_*}/c_{v_*}$ & $\gamma = O(1)$\\
	    $ \Theta$ & $T_*$&&& \\
	    $P$ & $p_*$& & & \\
		\end{tabular}    
	\end{center}
	\caption{Dimensionless variables and parameters, as in the previous analyses of \cite{chini_2014} and \cite{michel_chini_2019} except for the aspect ratio, the Reynolds number and the P\'eclet number, which in the present work are $O(1)$ quantities (asymptotically).\\} 

\label{tab:scalings}
\end{table}

\subsection{Asymptotic analysis}
A multiple time-scale analysis of the governing equations is performed (i) to disentangle the dynamics of the fast acoustic waves from the comparatively slowly evolving streaming flow and (ii) to consistently suppress negligible terms to simplify the numerical implementation and facilitate appropriate physical interpretation. To this end, we expand the various fields in powers of the small dimensionless parameter $\epsilon$:
\begin{eqnarray}
    (u,v) &=& \epsilon (u_1,v_1) + O(\epsilon^2),\\ 
    P &=& 1 + \epsilon \pi_1 + \epsilon^2 \pi_2 +  O(\epsilon^3),\\
    \Theta &=& 1 + \Gamma y + \Theta_0 + \epsilon \Theta_1 + O(\epsilon^2),\label{eqn:Texpansion}\\
    \rho &=& \rho_0 + \epsilon \rho_1 + O(\epsilon^2).
\end{eqnarray}
Note that the dimensionless temperature $\tilde{T}/T_*$ is denoted by $\Theta$, cf.~(\ref{eqn:Texpansion}), rather than by $T$, the latter notation being reserved for the `slow' time variable (see below).

In absence of acoustic waves, $u=v=0$ and, in steady state, the temperature profile reduces to the conduction profile, $\Theta=1+\Gamma y$, with the pressure being uniform since gravity is neglected.  When excited, the leading-order acoustic velocity is $O(\epsilon)$, by definition since $U_* = \epsilon a_*$, and similarly small perturbations in the pressure and temperature fields are generated. The $O(1)$ temperature disturbance $\Theta_0$ arises from the reorganization of temperature inhomogeneities by the strong streaming flow, as shall be made explicit in the wave/mean-flow interaction equations given subsequently.

The distinction between the waves and the streaming flow can be made based on a separation of time scales. While the acoustic wave fields exhibit fast oscillations with zero mean value, the streaming flow effectively remains constant over this time scale, evolving instead on a slow time $T \equiv \epsilon t$. 
The formal separation is captured via a WKBJ approximation through the introduction of a rapidly varying phase $\phi(t) = \Phi(T)/\epsilon$. In this framework, any field $f(x,y,t)$ is expressed as $f(x,y,\phi,T)$, where $\phi$ and $T$ are taken to be independent variables, and thus can be decomposed as
\begin{equation}
    f(x,y,\phi,T ) = \bar{f}(x,y,T) + f'(x,y,\phi;T).
\end{equation}
Here, the streaming flow component is represented by 
\begin{equation}
    \bar{f}(x,y,T) = \frac{1}{2\pi n} \int_\phi^{\phi + 2 n \pi} f(x,y,s,T) ds
\end{equation}
for sufficiently large positive integer $n$ (and arbitrary $\phi$). In contrast, $f'(x,y,\phi ;T)$ describes the acoustic wave of zero mean value ($\overline{f'}(x,y,\phi;T)  = 0$). Finally, the instantaneous angular frequency $\omega(T)$ satisfies

\begin{equation}
    \omega(T) = \frac{\mathrm{d}\phi}{\mathrm{d}t} = \frac{\mathrm{d}\Phi}{\mathrm{d}T} = \omega_0(T) + O(\epsilon).
\end{equation}

\subsection{Leading-order multiscale wave/mean-flow interaction equations}
\label{sim_equations}
This multiscale analytical approach has been used by \cite{chini_2014} to derive a closed set of equations describing the coupled dynamics of the acoustic wave and the streaming flow. The present formulation differs only in the scaling of the aspect ratio and diffusive terms, which can be easily traced in the derivation. Here, we simply state the resulting leading-order equations. 

The dynamics of the streaming flow is governed by the following mean-flow system:
\begin{align}
        \bar{\rho}_0 \left(  \partial_T \bar{u}_1 + \bar{u}_1 \partial_x \bar{u}_1 + \bar{v}_1 \partial_y \bar{u}_1   \right) = & - \frac{\partial_x \bar{\pi}_2 }{\gamma} - \left[ \partial_x \left(\bar{\rho}_0 \overline{u_1^{'2}} \right) + \partial_y \left(\bar{\rho}_0 \overline{u_1'v_1'} \right) \right] \nonumber \\
        & + \frac{1}{Re} \left[ \partial_{xx} \bar{u}_1 + \frac{1}{\delta^2}  \partial_{yy} \bar{u}_1 + \frac{1}{3} \left( \partial_{xx} \bar{u}_1 + \partial_{xy} \bar{v}_1 \right) \right],
        \label{str_eq1}
\end{align}

\begin{align}
        \bar{\rho}_0 \left(  \partial_T \bar{v}_1 + \bar{u}_1 \partial_x \bar{v}_1 + \bar{v}_1 \partial_y \bar{v}_1   \right) = & - \frac{\partial_y \bar{\pi}_2 }{\gamma \delta^2} - \left[ \partial_x \left(\bar{\rho}_0 \overline{u_1' v_1'} \right) + \partial_y \left(\bar{\rho}_0 \overline{v_1^{'2}} \right) \right] \nonumber \\
        & + \frac{1}{Re} \left[ \partial_{xx} \bar{v}_1 + \frac{1}{\delta^2}  \partial_{yy} \bar{v}_1 + \frac{1}{3\delta^2} \left( \partial_{xy} \bar{u}_1 +  \partial_{yy} \bar{v}_1 \right) \right],
        \label{str_eq2}
\end{align}

\begin{equation}
    \partial_T \bar{\rho}_0 + \partial_x \left( \bar{\rho}_0 \bar{u}_1 \right) + \partial_y \left( \bar{\rho}_0 \bar{v}_1 \right) = 0,
    \label{str_eq3}
\end{equation}

\begin{align}
 \bar{\rho}_0 \left[   \partial_T \bar{\Theta}_0 + \bar{u}_1 \partial_x \bar{\Theta}_0 + \bar{v}_1 \left(\Gamma + \partial_y \bar{\Theta}_0 \right) \right] =& (1-\gamma) \left(\partial_x \bar{u}_1 + \partial_y \bar{v}_1 \right) \nonumber \\
        &+ \frac{\gamma}{Pe} \left( \partial_{xx} \bar{\Theta}_0 + \frac{1}{\delta^2} \partial_{yy} \bar{\Theta}_0 \right),
        \label{str_eq4}
\end{align}

\begin{equation}
    \bar{\rho}_0 = \frac{1}{1+\Gamma y + \bar{\Theta}_0}.
    \label{str_eq5}
\end{equation}
This system describes the wave-averaged motion of a compressible ideal gas driven by an acoustic force density \blu{$\textbf{f}_\mathrm{ac}=- \nabla \cdot (\overline{\rho}_0 \overline{\mathbf{u}_1'\mathbf{u}_1'})$ that, for an ideal gas, can be reduced to $-(1/2) \overline{|\mathbf{u}_1'|^2} \nabla \overline{\rho}_0$ (up to a pressure gradient), as reported in \eqref{RSD_Karlsen} \citep{Karlsen_2016}}. If the acoustic force density is fixed, the problem reduces to that of forced convection. 

In baroclinic acoustic streaming, however, the situation is considerably more subtle, as the slowly evolving temperature disturbance $\bar{\Theta}_0$, and hence the density field $\bar{\rho}_0$, feeds back on the acoustic wave. This coupling is clearly evident in the corresponding equations for the waves, \emph{viz.}
\begin{equation}
\omega_0 \bar{\rho}_0 \partial_\phi u_1' + \frac{1}{\gamma} \partial_x \pi_1' = 0,
\label{ac_eq1}
\end{equation}

\begin{equation}
\omega_0 \bar{\rho}_0 \partial_\phi v_1' + \frac{1}{\gamma \delta^2} \partial_y \pi_1' = 0,
\label{ac_eq2}
\end{equation}

\begin{equation}
\omega_0  \partial_\phi \rho_1' + \partial_x \left( \bar{\rho}_0  u_1' \right) + \partial_y \left( \bar{\rho}_0  v_1'\right) = 0,
\label{ac_eq3}
\end{equation}

\begin{equation}
    \omega_0 \partial_\phi \Theta_1' + u_1' \partial_x \bar{\Theta}_0 + v_1' \left(\Gamma + \partial_y \bar{\Theta}_0  \right) +  \frac{\gamma -1}{\bar{\rho}_0 } \left( \partial_x u_1' + \partial_y v_1' \right)=0,  
    \label{ac_eq4}
\end{equation}

\begin{equation}
    \pi_1' = \frac{\rho_1'}{\bar{\rho}_0 } + \bar{\rho}_0 \Theta_1'.
    \label{ac_eq5}
\end{equation}
On the fast time scale, the dynamics of the acoustic waves is governed by a linear homogeneous system, the solution of which consists of a sum of modes of various amplitudes and angular frequencies. Here, however, we assume that only one acoustic mode is externally forced, i.e. only one wave has finite amplitude. This acoustic mode has dimensional horizontal wavenumber $k_*$ and is the solution of the eigenvalue problem for the wave associated with the lowest eigenvalue (i.e. the smallest angular frequency) that also has a non-uniform horizontal structure. Thus, any oscillatory field $f_1'$, representing $(u_1', v_1', \pi_1', \Theta_1', \rho_1')$, can be expressed as 
\begin{equation}
    f_1'(x,y,\phi ;T) = \frac{A(T)}{2}\left[\hat{f}_1(x,y,T) e^{i\:\phi} + \mathrm{c.c.} \right],
\end{equation}
where $\mathrm{c.c.}$ denotes the complex conjugate, $A(T)$ is the amplitude of the mode and $\hat{f}_1$ is a complex function that characterizes its spatial structure. For $A(T)$ to be uniquely specified, a normalization condition must be imposed; we opt to require the eigenmodes to satisfy 
\begin{equation}
\underset{x\in[0,2\pi], y\in [0,1]}{\mathrm{max}} \left \vert \hat{\pi}_1(x,y,T)\right \vert = 1.   
\end{equation}
\blu{In practice, this amplitude $A$ can be readily inferred experimentally from the (dimensional) steady-state amplitude of the pressure oscillations measured by a sensor at $(x_s,y_s)$, which yields $\epsilon A \vert \hat{\pi}(x_s,y_s)\vert p_*$.}
In the limit of very narrow channels \citep{michel_chini_2019}, the  eigenvalue problem for the lowest mode can be reduced to a one-dimensional problem (in $x$) that can be solved analytically for a linear temperature profile. Moreover, in that limit, the multiple scale analysis can be carried out to next order to obtain an equation governing the temporal evolution of the modal amplitude $A(T)$. Unfortunately, the corresponding higher-order analysis cannot be easily executed for the fully two-dimensional eigenvalue problem obtained here. The amplitude of the acoustic mode therefore will be assumed to be constant, i.e. an external control parameter, in this investigation for simplicity. Experimentally, this specification could be achieved via slowly-varying acoustic forcing mechanisms controlled by the feedback of a pressure sensor.

\section{Numerical Simulations of the Wave/Mean-Flow Equations}
\label{sec:results}
\subsection{Methods}
\label{sec:methods}
\FloatBarrier

The two time-scale quasilinear wave/mean-flow interaction system obtained in \S \ref{sim_equations} is integrated numerically over the slow time variable $T$. Specifically, we solve the initial-value problem (\ref{str_eq1})--(\ref{str_eq5}) governing the slow evolution of the streaming fields using a third-order four-stage Runge-Kutta scheme implemented in the spectral computational framework \emph{Dedalus} \citep{dedalus}. Equations (\ref{str_eq3}) and (\ref{str_eq4}) are reformulated to yield a divergence condition on the streaming velocity field and a modified energy equation, \emph{viz.}
\begin{equation}
    \partial_x \bar{u}_1 + \partial_y \bar{v}_1 = \frac{1}{Pe} \left( \partial_{xx} \bar{\Theta}_0 + \frac{1}{\delta^2} \partial_{yy} \bar{\Theta}_0 \right),
\label{str_eq6}
\end{equation}
\begin{align}
 \bar{\rho}_0 \left[   \partial_T \bar{\Theta}_0 + \bar{u}_1 \partial_x \bar{\Theta}_0 + \bar{v}_1 \left(\Gamma + \partial_y \bar{\Theta}_0 \right) \right] = \frac{1}{Pe} \left( \partial_{xx} \bar{\Theta}_0 + \frac{1}{\delta^2} \partial_{yy} \bar{\Theta}_0 \right).
\label{str_eq7}
\end{align}
At each (slow) time step, the acoustic fields $\hat{u}_1$ and $\hat{v}_1$ are required to evaluate the wave-induced Reynolds stress divergence. Accordingly, the set of equations (\ref{ac_eq1})--(\ref{ac_eq5}) governing the fast wave dynamics is consolidated to form \blu{an} eigenvalue problem for the fluctuating pressure field: 
\begin{equation}
    \blu{\partial_{x}\left[\overline{\rho}_0^{-1}  \partial_{x} \pi_1'\right] + \delta^{-2}\partial_{y}\left[ \overline{\rho}_0^{-1} \partial_{y} \pi_1'\right] = \omega_{0}^2\:\partial_{\phi \phi} \pi_1'.}
    \label{ac_eq6}
\end{equation}
We solve, \blu{at each time step, the 2D eigenvalue problem} (\ref{ac_eq6}) in Matlab using a Fourier--Chebyshev collocation discretization \citep{Trefethen, Driscoll2014}.
In both the Dedalus and Matlab codes, Fourier series in the periodic $x$-direction and Chebyshev polynomials in the wall-normal $y$-direction are used to represent all field variables. The number of modes is chosen to correspond to a $122 \times 122$ spatial grid resolution, and the time step $\Delta T=0.01$. Numerical convergence in space and time has been verified.

To model a physical experiment in which a specific acoustic mode is forced for time $T \geqslant 0$, the steady basic conduction state ($\bar{\Theta}_0 = \bar{v}_0=\bar{u}_0=0$) is taken as the initial condition. \blu{The solution of the eigenvalue problem targets the horizontal standing wave of smallest non-zero eigenvalue. For particular values of $\delta$, standing waves along the vertical direction may share the same angular frequency as the horizontal mode of interest, but in practice the nature of the forcing (e.g. oscillating horizontal or vertical walls) would enable only one of these modes to be excited. Both the resulting eigenfunction and eigenfrequency can vary with $T$.} The results discussed in this section are obtained for various acoustic wave amplitudes $A$ and aspect ratios $\delta$, keeping the following parameters fixed:
\begin{equation}
    \Gamma = 0.3, \quad\quad \gamma = 1.4, \quad\quad Re = 2500, \quad\quad Pe = 1775.\label{DIM_param}
\end{equation}
This set of parameters corresponds to using air as the working fluid, for which $\gamma = 1.4$ and the Prandtl number $Pe/Re=0.71$, inside an acoustic cavity similar to the one employed by \citet{Michel_2021} but with \blu{larger Reynolds and P\'eclet numbers to clearly reveal strongly nonlinear dynamical behavior}. We monitor the top- and bottom-wall Nusselt numbers, defined as the ratio of the respective heat fluxes at the upper and lower channel walls to the strictly conductive heat flux and evaluated as 
\begin{equation}
    Nu_\mathrm{t} (T)=1+ \frac{1}{2\pi \Gamma} \int_0^{2\pi} \partial_y \Bar{\Theta}_0(x,1,T) ~dx, ~     Nu_\mathrm{b} (T)=1+ \frac{1}{2\pi \Gamma} \int_0^{2\pi} \partial_y \Bar{\Theta}_0(x,0,T) ~dx,
\end{equation}
where the subscripts `t' and `b' denote `top' and `bottom'.
In a steady state, integrating (\ref{str_eq6}) over the entire domain yields the expected equality $Nu_\mathrm{t}(\infty)=Nu_\mathrm{b}(\infty)\equiv Nu$.

This numerical approach, in which an eigenvalue problem is solved at each time step to account for the slow evolution of the acoustic mode, has been introduced in \cite{Karlsen_2018} and \cite{michel_chini_2019} \blu{and is necessary to consistently capture the acoustic force density $-(1/2) \overline{|\mathbf{u}_1'|^2} \nabla \overline{\rho}_0$, since $\mathbf{u}_1'$ depends non-locally on the entire background density field $\overline{\rho}_0(x,y,T)$}. Nevertheless, the relevance of regularly updating the acoustic wave fields remains insufficiently documented, leading to its arbitrary omission in most subsequent investigations. To quantitatively assess the importance of two-way coupling, we therefore also performed a set of numerical simulations, referred to as `one-way coupled', in which the temporal integration is carried out without solving the eigenvalue problem at each time step. \blu{Instead, $\mathbf{f}_{ac}(x,y,T)$ is set to $-(1/2) \overline{|\mathbf{u}_1'|^2} \nabla \overline{\rho}_0$ evaluated at initial time $T=0$. We acknowledge that several alternative ``one-way coupling" approximations can be proposed, such as (i) only fixing $\overline{|\mathbf{u}_1'|^2}$ in time and evaluating $\mathbf{f}_{ac}$ using the actual density gradient $\nabla \overline{\rho}_0$ or (ii) setting the fluctuating pressure 
$\pi'(x,y,T,\phi) = A \cos(x) \cos(\phi)$ 
(i.e. a single Fourier mode in $x$ that is independent of $y$)
and evaluating the acoustic force density with the fluctuating acoustic velocity obtained from \eqref{ac_eq1} and the current density gradient. The latter option
may be appropriate for
well-mixed steady-states realised at large dimensionless amplitude $A$ (see \S \ref{S_Evolution_Nu}).}


\subsection{Convergence to a steady state}
\label{sec:convergence}

For small and intermediate acoustic wave amplitudes $A$, our numerical simulations reveal that the system converges to parameter-dependent steady states. A typical run for $\delta=4$ and $A=4$ shows that a few hundred slow time units are required to reach this steady state; see figure~\ref{NuvsT} and the animation included in the supplemental material. This requirement highlights the importance of using a multiple-scale algorithm, since a single time unit corresponds in practice to hundreds or thousands of acoustic cycles (cf. $\epsilon \approx 10^{-4}$ for the setup of \cite{Michel_2021} discussed in \S \ref{Scaling_ND}). During the transient regime, the top and bottom heat fluxes evolve non-monotonically, ultimately converging to a common value significantly larger than the conductive heat flux. The variation of $Nu$ with the amplitude of the wave $A$ and with the aspect ratio $\delta$ will be discussed in \S \ref{S_Evolution_Nu}.


\begin{figure}
  \centerline{\includegraphics[width=1\linewidth]{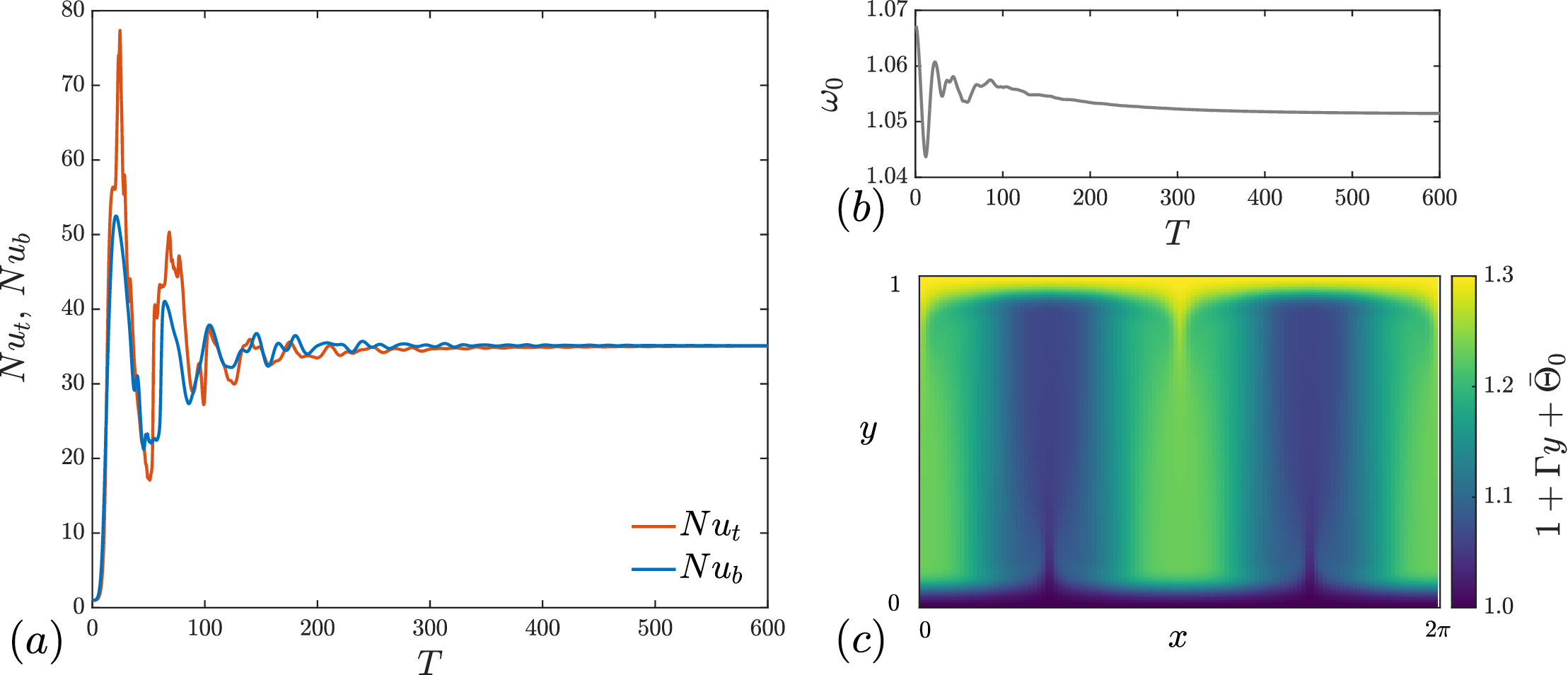}}
  \caption{Time series for $A=4$ and $\delta=4$ of (a)~the top and bottom Nusselt numbers $(Nu_\mathrm{t},Nu_\mathrm{b})$ and (b) of the acoustic-wave angular frequency $\omega_{0}$. The total steady-state temperature field $1+ \Gamma y + \overline{\Theta}_{0}$ shown in (c) exhibits strong variations in $x$ associated with localized jets at $x=\lbrace 0,\pi/2, \pi, 3\pi/2, 2\pi \rbrace$ and boundary layers in $y$ close to both walls. 
  }
\label{NuvsT}
\end{figure}

This steady-state exhibits both thermal boundary layers and vertical jets. Low-amplitude waves ($A\ll 1$) generate smooth cellular structures (not shown here, for brevity), similar to the ones computed theoretically and numerically in \cite{michel_chini_2019} for $\delta \ll 1$ and in the direct numerical simulations of \cite{LIN_2008,AKTAS_2010, Baran_2022, Malecha_2023}. 

\begin{figure}
  \centerline{\includegraphics[width=1\linewidth]{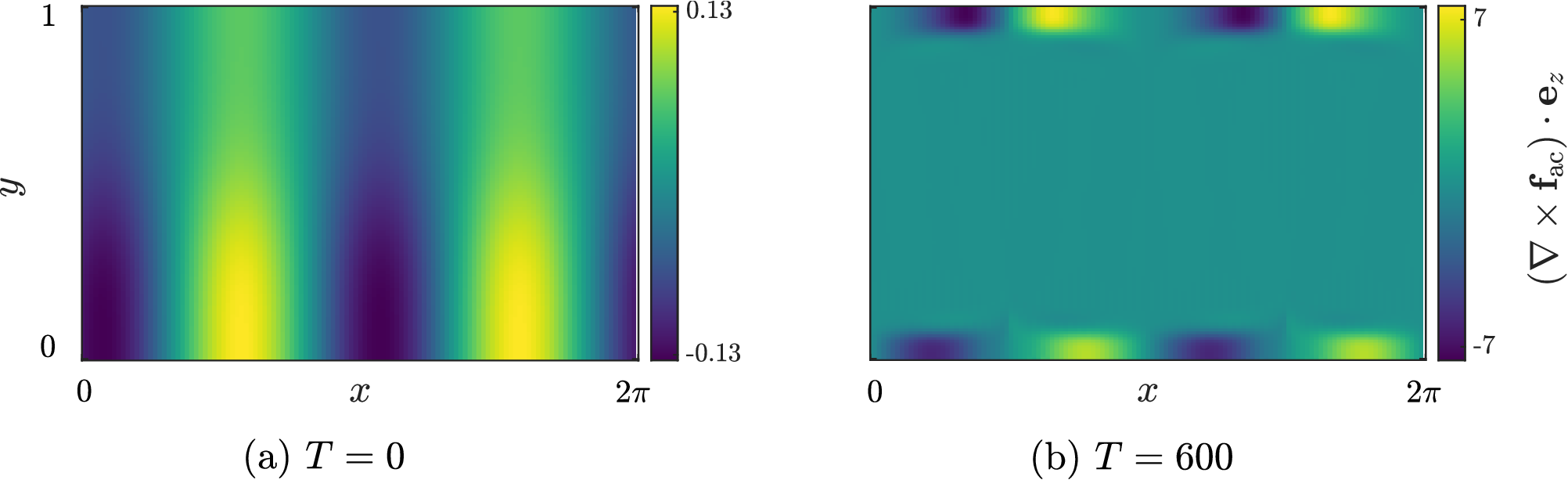}}
  \caption{Evolution for $A=4$ and $\delta = 4$ of the curl of the acoustic force density $\nabla \times \mathbf{f}_\mathrm{ac}$: the initial condition (a) and the steady-state (b). $\mathbf{e}_{z} \equiv \mathbf{e}_{x}\times\mathbf{e}_{y}$, where $\mathbf{e}_{x}$ and $\mathbf{e}_{y}$ are unit vectors in the $x$ and $y$ directions, respectively. The feedback from the evolving streaming density field $\overline{\rho}_{0}$ leads to the localisation of $\nabla \times \mathbf{f}_\mathrm{ac}$ near the upper and lower walls.
  }
\label{delxf}
\end{figure}

To gain a more detailed understanding of these steady states, we further analyse the Reynolds stress divergence, also referred to as the acoustic force density $\mathbf{f}_\mathrm{ac}$, resulting from the acoustic wave. This force density, corresponding to the second terms on the right-hand sides of (\ref{str_eq1})--(\ref{str_eq2}), is balanced by mean inertia, mean viscous forces and the mean pressure gradient. Crucially, the part of this wave forcing that actually sustains the streaming flow, i.e., the part that is not balanced by a mean pressure gradient, is given by the curl $\nabla \times \mathbf{f}_\mathrm{ac}$. In figure~\ref{delxf}, we plot this (scalar) field at the initial time, when the mean temperature distribution corresponds to the linear conduction profile, and in the steady state. The striking difference confirms the two-way coupling between the waves and the streaming flow in this system, in contrast to the usual forced convection configuration. A streaming flow generated by the curl of the acoustic force modifies the mean density distribution and, hence, the properties of the acoustic waves and therefore also $\nabla \times \mathbf{f}_\mathrm{ac}$.

\begin{figure}
  \centerline{\includegraphics[width=1\linewidth]{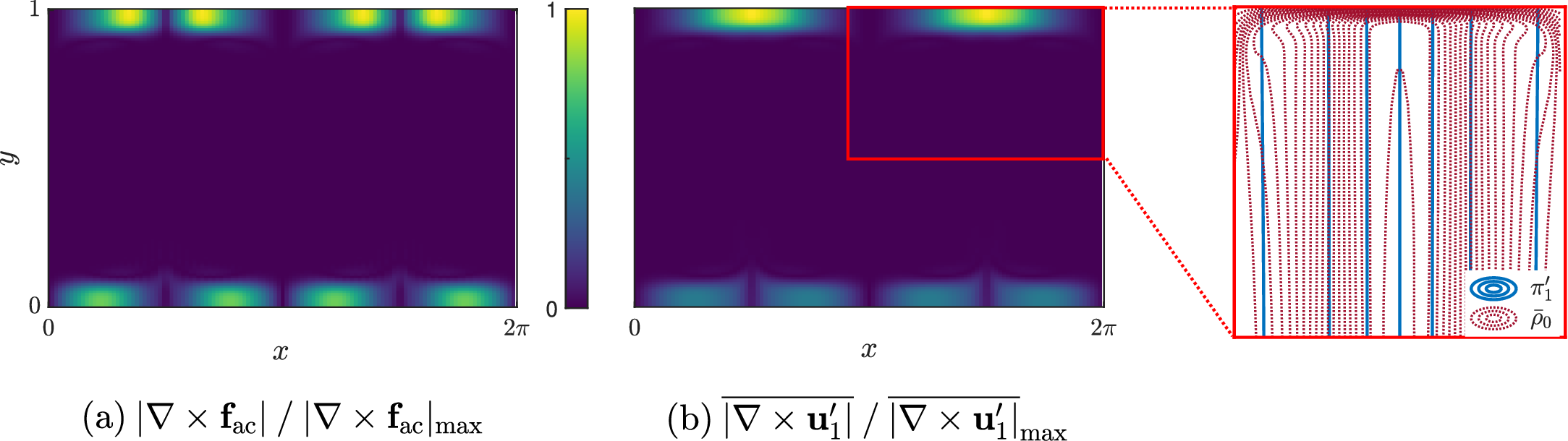}}
  \caption{\blu{Comparison for $A=4$ and $\delta = 4$ of the normalized steady-state amplitudes of (a) the curl of the Reynolds stress divergence $\nabla\times \mathbf{f}_\mathrm{ac}$ and (b) the acoustic-wave vorticity $\nabla\times \mathbf{u}_1'$. Since a \emph{horizontal} standing acoustic mode is considered, the fluctuating isobars (isovalues of $p'$) are essentially vertical (see inset in (b)) and acoustic vorticity is therefore localized where vertical gradients of density exist (see \eqref{vorticity}), i.e. in the top and bottom boundary layers.} 
  }
\label{fac_vort}
\end{figure}

\blu{As noted in \S\ref{sec:intro}, the generation of fluctuating vorticity in a gas is key to understanding streaming flows. In a \textit{homogeneous} gas, a nonzero acoustic-wave vorticity implies that the acoustic force density
\begin{equation}
    \mathbf{f}_{ac}  \equiv - \nabla \cdot \left(\overline{\rho}_0\,\overline{\mathbf{u}_1'\mathbf{u}_1'}\right)= -\overline{\rho}_0\, \nabla \cdot \left(\overline{\mathbf{u}_1'\mathbf{u}_1'}\right)=  \text{gradient term } - \overline{\rho}_0\,\overline{\left(\nabla \times \mathbf{u}_1'\right) \times \mathbf{u}_1'}
\end{equation}
generically cannot be balanced by a gradient term and, thus, may drive a streaming flow. Acoustic-wave vorticity in that case is usually generated by thermal and/or viscous diffusion in thin oscillating boundary layers. In an \textit{inhomogeneous and inviscid} gas, the acoustic force density instead reduces to \citep{Karlsen_2016}
\begin{align}
        \mathbf{f}_{ac} &\equiv - \nabla \cdot \left(\overline{\rho}_0\,\overline{\mathbf{u}_1'\mathbf{u}_1'}\right)= \text{gradient term } - \frac{1}{2}\,\overline{\left\vert \mathbf{u}_1' \right\vert^2}\,\nabla \overline{\rho}_{0}\label{vorticity_e1}\\&= \text{gradient term } - \frac{1}{2}\left(\overline{\rho}_0\,\overline{\left(\nabla \times \mathbf{u}_1'\right) \times \mathbf{u}_1'}  + \overline{( \mathbf{u}_1' \cdot \nabla \overline{\rho}_0) \mathbf{u}_1'}   \right),     \label{vorticity_e2}  
\end{align} 
where \eqref{vorticity_e2} can be directly derived from \eqref{vorticity_e1} and the leading order fluctuating equations.} \blu{In the absence of acoustic-wave vorticity, i.e. in the absence of fluctuating baroclinicity $\nabla\overline{\rho}_0 \times \nabla \pi_1' = 0$,
the steady state contribution to $\nabla\times\mathbf{f}_{ac}$ from the term $\overline{(\mathbf{u}_1' \cdot \nabla \overline{\rho}_0) \mathbf{u}_1'}$ in \eqref{vorticity_e2} is negligible. Thus, even in \textit{inhomogeneous} gases, acoustic-wave vorticity plays a crucial role in driving streaming flows.} 
Physical insights follows from and qualitative predictions can be made based on this understanding. First, it accounts for the localization of $\nabla \times \mathbf{f}_\mathrm{ac}$ close to the walls, as evident in figure~\ref{delxf}. As shown in figure~\ref{fac_vort}, $\nabla \times \mathbf{f}_\mathrm{ac}$ is significant only in regions where acoustic-wave vorticity also is large in magnitude; and wave vorticity is confined to thermal boundary layers, since the fluctuating isobars and the mean isopycnals are almost orthogonal there (see \eqref{vorticity}). Secondly, this understanding explains why the \blu{orientation} of the standing acoustic wave is crucial in such systems. Consider, for example, a configuration in which the initial conduction state of the present setup is perturbed with a standing wave oscillating along the \textit{vertical} rather than horizontal direction: the fluctuating isobars and the mean isopycnals would then be approximately aligned, resulting in zero acoustic vorticity and, consequently, negligible baroclinic acoustic streaming (see also \citet{Kumar_2021}). 


\subsection{Dependence on wave amplitude}
\label{sec:Amp_dependence}

\begin{figure}
  \centerline{\includegraphics[width=1\linewidth]{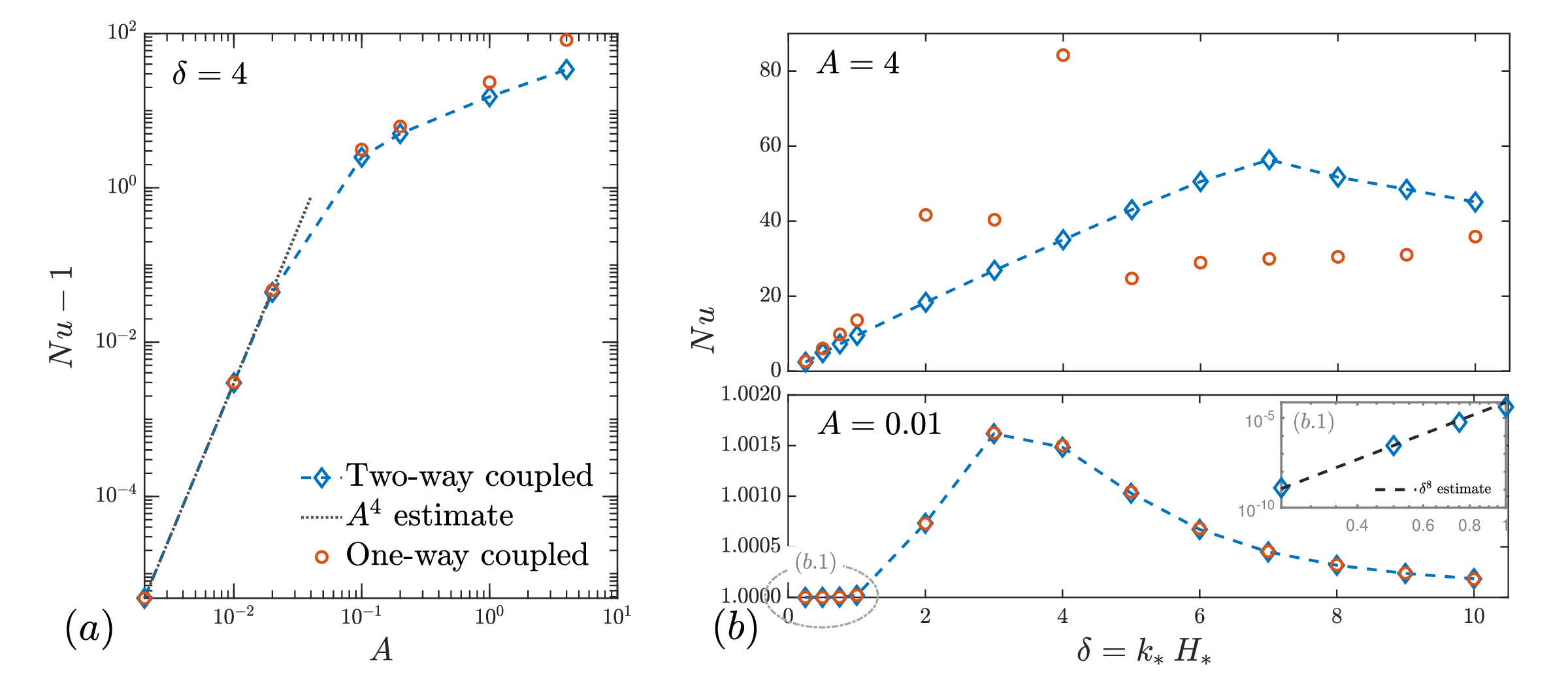}}
  \caption{\blu{(a) Steady-state Nusselt number $Nu$ (minus one) as a function of the acoustic wave amplitude $A$ for $\delta = 4$, along with the asymptotic prediction $Nu-1 \propto A^4$ that can be derived in the limit $A \ll 1$. (b) Steady-state Nusselt number $Nu$ as a function of the aspect ratio $\delta$ for both $A=4$ and $A=0.01$. The one-way coupled simulations, in which the evolution of the acoustic waves is neglected, provide accurate results only in the small amplitude limit $A \ll 1$.} }
\label{Nuvsdelta}
\end{figure}

The influence of the acoustic wave amplitude $A$ is investigated through a suite of numerical simulations performed for the same fixed aspect ratio $\delta = 4$ but for various values of $A \in \left[2 \times 10^{-3}, 4\right]$. All simulations converge to steady states, with the Nusselt number $Nu$ plotted in figure~\ref{Nuvsdelta} (left). The results are compared to one-way coupled simulations (defined in \S \ref{sec:methods}) that also converge to steady states for $A \in \left[2 \times 10^{-3}, 0.2\right]$. In the range $A \in \left[1, 4\right]$, the one-way coupled dynamics evolves to \emph{statistically} stationary states, and the corresponding Nusselt numbers are obtained by time-averaging over more than $10^3$ slow time units.

In the limit of small acoustic-wave amplitude $A \ll 1$, the acoustic force density $\mathbf{f}_\mathrm{ac}\propto A^2$ generates weak streaming flows and correspondingly small changes in the background temperature fields. Although of limited practical interest since the resulting Nusselt numbers are close to unity, this regime can be easily analysed theoretically by neglecting the evolution of the acoustic wave properties, i.e., by assuming that the one-way coupled dynamics provides an accurate approximation. Similarly to the analysis of \cite{michel_chini_2019}, which assumes $\delta \ll 1$, we obtain for $\delta = O(1)$ that
\begin{equation}
Nu -1 \underset{A \ll 1}{ = } A^4 Re^2 Pe^2 \gamma^{-4} F(\delta,\Gamma), \label{1W_Nu}
\end{equation}
with $F(\delta,\Gamma)$ a function that has to be computed numerically; in the limit $\delta\ll 1$, $F(\delta,\Gamma) = \delta^8  G(\Gamma)$ for some function $G(\Gamma)$, as predicted by \citet{michel_chini_2019}. The results of the numerical simulations reported in figure~\ref{Nuvsdelta} support the $A^4$ scaling for $A \leqslant 0.02$, a range for which one-way and two-way coupled simulations yield nearly indistinguishable results.

Figure~\ref{Nuvsdelta} also reveals that the one-way coupled numerical simulations and resulting theoretical prediction \eqref{1W_Nu} fail to capture the two-way coupling that emerges for $A \geqslant 0.1$. In this regime, $Nu$ is significantly larger than unity, and accurate results can be obtained only by using the full machinery of the multiple-scale numerical algorithm implemented in our numerical simulations.

\subsection{Dependence on aspect ratio}
\label{S_Evolution_Nu}

The impact of varying the aspect ratio $\delta  = k_* H_*$ is investigated with a set of numerical simulations performed for fixed $A = 0.01$ or 4 and varying $\delta \in \left[ 0.25, 10 \right]$. This variation can be interpreted as being achieved, for example, by adjusting the channel height $H_*$, since $\delta$ is the only parameter involving $H_*$ (see table~\ref{tab:scalings}).

The Nusselt numbers reported in figure~\ref{Nuvsdelta}$b$ for small acoustic amplitude $A=0.01$ reach a maximum for $\delta \simeq 3$. Since the acoustic amplitude $A \ll 1$, the evolution of the acoustic wave properties can be neglected, and the one-way coupled simulations accurately describe this regime.  The Nusselt number is found to converge to unity as $\delta \rightarrow 0$, 
following the asymptotic scaling law $Nu-1 \propto \delta^8$ derived in \cite{michel_chini_2019}, 
which evidently here holds
for $\delta \leqslant 1$ \blu{(see the inset in figure~\ref{Nuvsdelta}$b$)}. For $A=4$, the feedback on the acoustic waves is essential, and two-way coupled simulations are required to reach a consistent steady state. In this case, the Nusselt number still exhibits significant variation with the aspect ratio, with a maximum value of $Nu = 56.4$ at $\delta = 7$. In this two-way coupled regime, the aspect ratio that maximizes the Nusselt number is expected to depend on all other parameters ($\Gamma$, $Re$, $Pe$, $\gamma$) in a complicated fashion. Given that, in practice, the height of the acoustic cavity is not easily modified once an experimental apparatus is built, these simulations are particularly valuable for identifying optimal parameters beforehand. 

\begin{figure}
  \centerline{\includegraphics[width=1\linewidth]{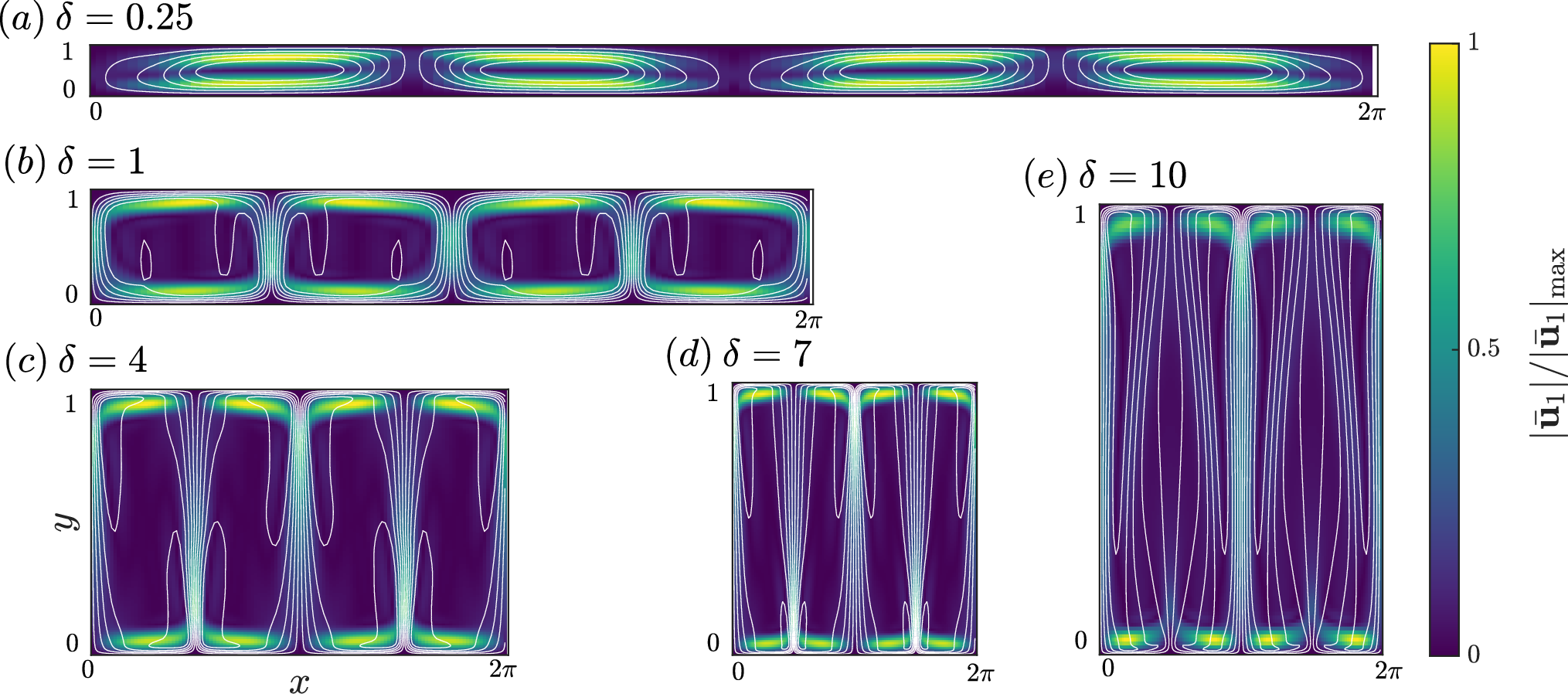}}
  \caption{Steady-state streaming velocity fields for $A=4$ and various aspect ratios $\delta$. The white lines are isovalues of the mass current potential $\phi_\rho$, defined such that $\overline{\rho}_0 \overline{\mathbf{u}}_1 = \nabla \phi_\rho$, and colors correspond to the normalised velocity magnitude $\left(|\overline{\mathbf{u}}_1|\:/\:|\overline{\mathbf{u}}_1|_\mathrm{max}\right)$. The ratio of the height to width of each panel is set to the respective value of $\delta$ to facilitate qualitative comparison. 
  }
\label{racetrack}
\end{figure}

\begin{figure}
  \centerline{\includegraphics[width=1\linewidth]{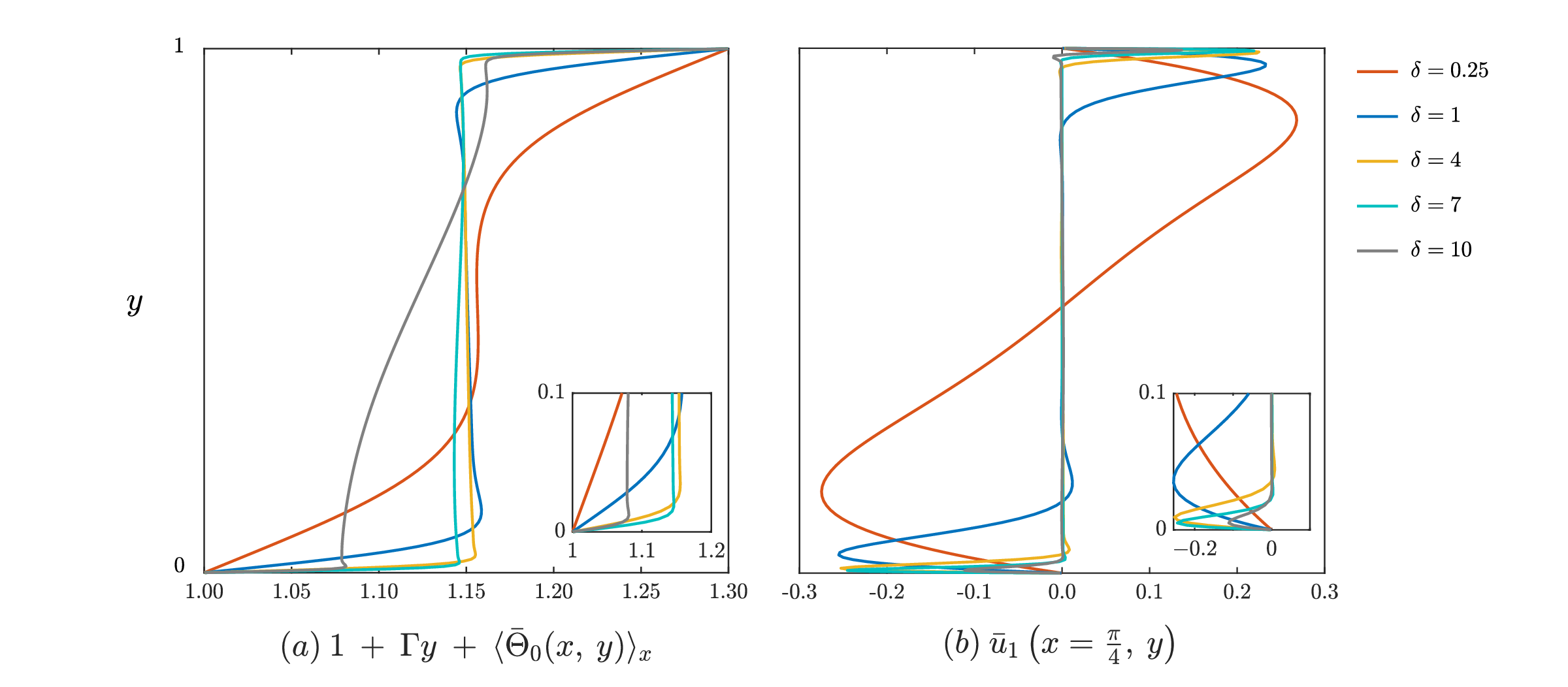}}
  \caption{Wall-normal profiles of steady-state streaming temperature and \blu{horizontal} velocity for $A=4$ and various aspect ratios $\delta$. (a) The total streaming temperature averaged over the horizontal $x$ direction, $1 +  \Gamma y + \langle \Bar{\Theta}_0(x,y)\rangle_x$. (b) The streaming $x$-velocity component at a fixed location $x = \pi/4$, i.e.  $\Bar{u}_1(x=\pi/4,y)$. The smooth profiles observed for $\delta = 0.25$ develop viscous and thermal boundary layers as $\delta$ increases.
}
\label{T_avg}
\end{figure}

The evolution of the streaming fields with $\delta$ is depicted in figures~\ref{racetrack} and~\ref{T_avg}. For the smallest value of $\delta$ ($=0.25$), diffusion in the wall-normal ($y$) direction prevails over inertia (see the factor $1/\delta^2$ in \eqref{str_eq1} and \eqref{str_eq2}) and smooth cellular structures are observed, similar to the ones reported in \cite{LIN_2008,AKTAS_2010, Baran_2022, Malecha_2023}. For larger values of $\delta$, top and bottom boundary layers accompanied by vertical jet-like structures are generated. For such flows, the $x$-averaged temperature $1+\Gamma y + \langle \Bar{\Theta}_0(x,y) \rangle_x$, where $\langle(\cdot)\rangle_x$ denotes an $x$-average, does not monotonically vary from 1 at $y=0$ to $1+ \Gamma$ at $y=1$, but instead displays thin regions of reversed temperature gradient. The same feature is observed in quasilinear models of buoyancy-driven convection \citep{Herring1968, OConnor_2021}. In these reduced models, convection is driven by the interaction between hydrodynamic modes, assumed to be linearised solutions that evolve on a fast time scale determined by an eigenvalue problem in which the more slowly-evolving temperature arises as a non-constant coefficient, and the slowly evolving temperature field itself, which is forced by flux divergences analogous to a wave-induced Reynolds stress divergence. Note that these emerging viscous and thermal boundary layers have a thickness that remains large compared to that of the oscillatory Stokes layers, induced by the no-slip condition at each wall, exhibited by the acoustic velocity field. These Stokes layers, which are passive in our analysis---only generating a higher-order (i.e. weaker) streaming flow---and, thus, self-consistently not resolved in our model, are of dimensional thickness $\delta_\mathrm{BL}^* = \sqrt{\mu_*/(\rho_*a_*k_*})$. The corresponding dimensionless expression is $\delta_\mathrm{BL}^* /H_* = \delta^{-1}\sqrt{\epsilon/Re}$.


\section{Conclusion}
\label{sec:conclusion}

The interaction between a thermally stratified gas and a standing acoustic wave \blu{has been} investigated in a channel with top and bottom walls of fixed but differing temperatures. As shown in \cite{chini_2014}, \cite{Karlsen_2016}, \cite{Karlsen_2018}  and \citet{michel_chini_2019}, a two-way coupling develops in such inhomogeneous systems between the streaming flow, which is forced by the standing acoustic wave, and the wave, whose properties are modified by the varying background temperature. The present study extends the work of \cite{michel_chini_2019} to characterize \blu{the resulting} heat fluxes in channels having heights comparable to the acoustic-wave wavelength. Moreover, the numerical simulations reported here highlight several features of baroclinic streaming that are not clearly apparent in the literature.

For the given geometry, the evolution of the Nusselt number $Nu$, i.e.  the heat flux normalized by that realized in the absence of acoustic wave, is of primary interest. In contrast with scaling laws derived in the limit of small acoustic amplitudes, see \eqref{1W_Nu}, and with previous investigations in narrow channels \citep{michel_chini_2019}, we report values of $Nu$ significantly larger than unity. This finding supports the practical potential of acoustics to enhance heat transfer in situations where forced convection is difficult to establish. The numerical algorithm that has been developed, coupling an initial value problem to an eigenvalue problem,  can be used to determine the geometry that maximizes the Nusselt number for given acoustic amplitudes and gas properties,  as shown for two specific sets of dimensionless parameters in figure~\ref{Nuvsdelta}$b$. Moreover, the results reported here constitute a reference case for future investigations seeking to further increase $Nu$ by tuning the boundary conditions (e.g. enforcing a constant heat flux instead of a constant temperature along the upper boundary to better model a heat source, or allowing for a non-zero wall-normal velocity to model a porous substrate). 

We also document several features of baroclinic streaming that are expected to be generic and apply, for instance, to the interaction of a standing acoustic wave with air surrounding a hot (or cold) cylinder or sphere or filling a half-plane above a warmer horizontal wall. In each scenario, these features can be traced to the baroclinic generation of acoustic-wave vorticity.  In strongly stratified gases, streaming flows are forced where the curl of the acoustic force density is non-zero; this only occurs where the acoustic-wave vorticity does not vanish, i.e. where the fluctuating isobars and mean isopycnals are approximately orthogonal. The following sequence of events therefore can be anticipated for an inhomogeneous gas suddenly forced by a standing acoustic wave. (i) Given a smooth initial distribution of the thermal inhomogeneity, acoustic-wave vorticity will spread throughout the domain and drive a streaming flow (figure~\ref{delxf}$a$). (ii) The streaming flow will mix the interior of the fluid, causing inhomogeneities in temperature to largely be confined to thin boundary layers (figures \ref{NuvsT}$c$ and \ref{T_avg}$a$). (iii) This evolution in the temperature field will modify the characteristics of the acoustic wave; in particular, acoustic vorticity, and hence the component of the acoustic force density effectively driving the streaming flow, also will localize in these boundary layers (figure~\ref{fac_vort}). (iv) The streaming flow will develop viscous boundary layers close to the hot (or cold) solid boundary (figure~\ref{T_avg}$b$) that results in jets sustaining the mixing in the interior of the domain (figure \ref{racetrack}$b$--$e$). \blu{We conjecture that a} transition to turbulence also may be expected to occur as the acoustic amplitude or Reynolds number is further increased.

\begin{figure}
  \centerline{\includegraphics[width=1\linewidth]{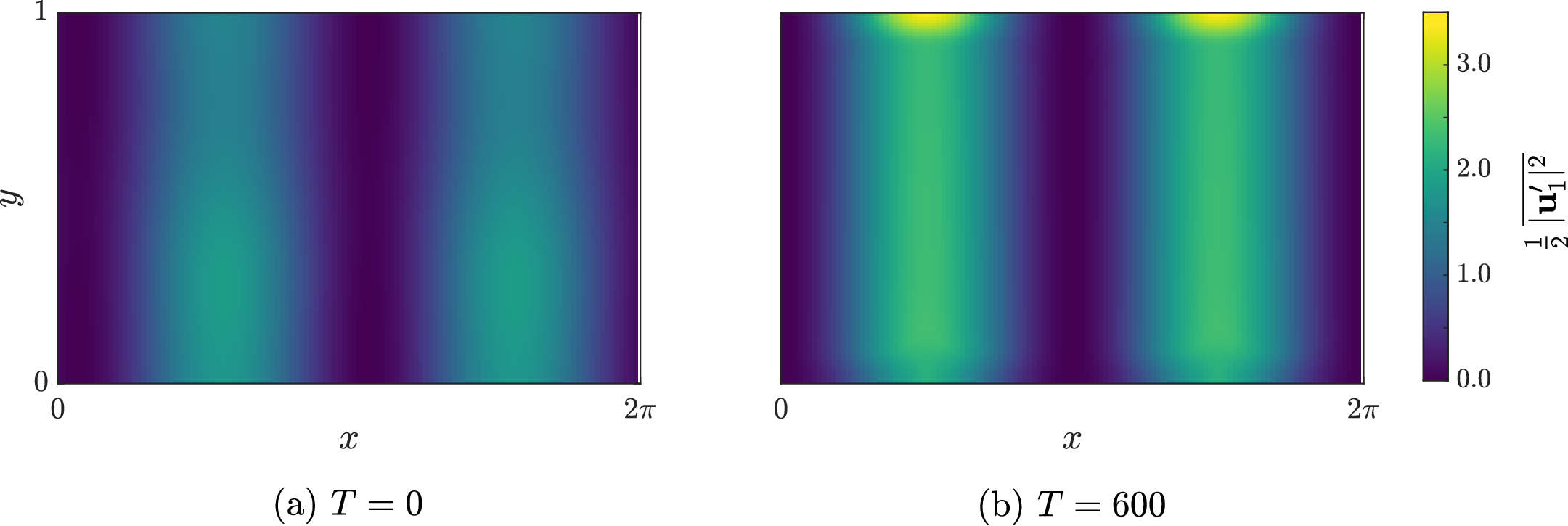}}
  \caption{Evolution for $A=4$ and $\delta = 4$ of the acoustic-wave kinetic energy $\frac{1}{2}\overline{|\mathbf{u}_1'|^2}$: the initial condition (a) and the steady-state (b). The evolution of the acoustic-wave kinetic energy, in addition to the streaming density field, also results in significant modifications to the acoustic force density \eqref{RSD_Karlsen}. 
  }
\label{ac_KE}
\end{figure}

This dynamical picture highlights the two-way coupling between the acoustic wave and streaming flow. To stress this point, we carried out one-way coupled simulations in which the acoustic wave fields are not updated as the streaming temperature field evolves. The resulting statistically stationary state is no longer strictly steady and the associated heat flux is not accurately estimated; see figure~\ref{Nuvsdelta}. Holding the acoustic velocities fixed in the evaluation of the acoustic force density \eqref{RSD_Karlsen}, either to simplify the numerical algorithm by not regularly solving an eigenvalue problem or to draw an analogy with an effective gravitational field \citep{koulakis_putterman_2021, Koulakis_2023}, therefore restricts the validity of the approach to waves of vanishing amplitude that do not drive any significant streaming flow and for which the advection of temperature inhomogeneities can be neglected.
Indeed, figure~\ref{ac_KE}, which shows the acoustic wave kinetic energy for both one-way ($a$) and properly two-way ($b$) coupled simulations, confirms that merely updating the mean density distribution in the acoustic force density \eqref{RSD_Karlsen} is \emph{not} sufficient for reliable and robust simulation of these flows.
Multiple scale analysis is the appropriate approach to treat such disparate time-scale coupling. Finally, this multiscale analysis can be extended to identify the scaling of the temperature differential across the domain $\Gamma = \Delta \Theta_*/T_*$ for which the feedback of the streaming flow on the wave must be included, i.e. to identify the crossover between pure Rayleigh streaming ($\Gamma = 0$) and baroclinic streaming ($\Gamma = O(1)$).\\

\paragraph{\textbf{Acknowledgements.} } This work is supported by CNES. We also acknowledge support from the Woods Hole Geophysical Fluid Dynamics Summer School (NSF 1829864), where this work was initiated.

\paragraph{\textbf{Declaration of Interests.}} The authors report no conflict of interest.\\

\bibliography{LASpaper1_rev1}

\end{document}